\documentclass[prl,superscriptaddress,twocolumn,showpacs,amsfonts,amsmath,floatfix]{revtex4}
\usepackage{graphicx}
\usepackage{bm}
\usepackage{amsmath}
\usepackage{hyperref}
\newcommand{\gmf}{$\gamma_{mf}$}
\newcommand{\gmean}{$\gamma_{\mu}$}
\newcommand{\dlg}{$\Delta{\gamma}$}

\begin{document}

\preprint{APS/123-SCN}

\title{Observation of Spatial Fluctuations of the Local Density of States in Random Media}

\author{M. D. Birowosuto}
\affiliation{Complex Photonic Systems, Faculty of Science and
Technology, and MESA+ Institute for Nanotechnology, University of Twente, P.O. Box 217, 7500 AE, Enschede,
The Netherlands}
\author{S. E. Skipetrov}
\affiliation{Laboratoire de Physique et Mod\'elisation des Milieux Condens\'es, Universit\'e Joseph Fourier, CNRS, 25 rue des Martyrs, BP 166, 38042 Grenoble, France}
\author{W. L. Vos}
\author{A. P. Mosk}
\email{A.P.Mosk@utwente.nl}
\affiliation{Complex Photonic Systems, Faculty of Science and
Technology, and MESA+ Institute for Nanotechnology, University of Twente, P.O. Box 217, 7500 AE, Enschede,
The Netherlands}

\date{\today}
\begin{abstract}
We experimentally study spatial fluctuations of the local density of states (LDOS) inside three-dimensional random photonic media. The LDOS is probed at many positions inside random photonic media by measuring emission rates of a large number of individual fluorescent nanospheres. The emission rates are observed to fluctuate spatially, and the variance of the fluctuations increases with the scattering strength. The measured variance of the emission rates agrees well with a model that takes into account the effect of the nearest scatterer only.
\end{abstract}

\pacs{42.25.Dd, 42.25.Fx, 42.50.Ct, 32.50.+d}
\keywords{fluctuations, local density of states (LDOS), time-resolved, random photonic media, intensity correlations.}

\maketitle

It is well-known that the spontaneous emission rate of an excited quantum emitter is not only a property of the emitter itself, but also depends on its surroundings on the nanoscale \cite{purcell,novotny}. Control of the emission rate of quantum emitters has been demonstrated with a wide range of systems, such as reflecting interfaces \cite{mirror}, microcavities \cite{cavity}, photonic crystals \cite{yablo,sprik,Ivan}, and plasmonic nanoantennae \cite{antenna}. The effect of the surrounding of the emitter is described by the local density of states (LDOS) that counts the number of optical modes available for emission at the position of the emitter \cite{novotny,sprik}.

In random photonic media that are promising systems to observe intriguing Anderson localization of light \cite{Wiersma}, it is an open question how spontaneous emission rates are affected by the surroundings. It has theoretically been predicted that the LDOS exhibits spatial fluctuations \cite{Mirlin,vanTiggelen,perez,perez2}. These LDOS fluctuations are determined by light scattering near the emitter and the variance of these fluctuations is of the order of the scattering strength ($1/k\ell$), where $k$ is the wave number of light in the medium and $\ell$ is the transport mean free path. In an infinite random medium the LDOS fluctuations are  essentially equivalent  to the  $C_{0}$  intensity correlation function \cite{vanTiggelen}. $C_{0}$ is an infinite-range correlation function for waves in random media \cite{shapiro}. The variance of LDOS fluctuations was also calculated using a single-scattering analytical model and was found to be highly sensitive to the level of scattering and absorption at the local scale \cite{perez}, as well as to the dipole orientation \cite{perez2}. No experimental observation of the LDOS fluctuations inside random photonic media has been reported so far.

In this Letter, we present an experimental study of fluctuations of the LDOS inside random photonic media. We use single nanoscale emitters as internal probes of the LDOS.
Time-resolved fluorescence is recorded for many emitters inside random media with scattering strengths $1/k\ell$ up to 0.12. The distribution of the spontaneous emission lifetimes is used to test theoretical predictions from several different models.
\begin{figure}[!ht]
\begin{center}
\includegraphics[width=3.25in]{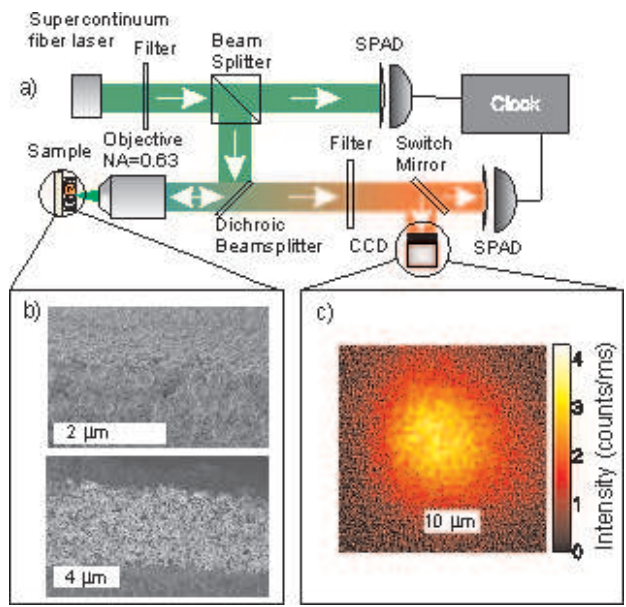}
\end{center}
\caption{\label{fig1}(color) (a) Time-correlated single-photon counting setup with modelocked supercontinuum fibre laser, CCD camera, and single-photon avalanche photodiode (SPAD). (b) Scanning electron micrographs (SEM) of random photonic media made of Polystyrene (PS) (top) and Zinc Oxide (ZnO) (bottom).
(c) Fluorescence image of a single fluorescent nanosphere with a diameter of 22 nm embedded in a 7.4(5) $\mu$m-thick slab of PS random photonic medium at a depth $d$ = 3.3 $\mu$m.}
\end{figure}

The experimental setup shown in Fig. \ref{fig1}(a) is designed to optically probe a single emitter located deep inside a random medium, and to measure its time-resolved emission. The emitters are optically excited at  a wavelength $\lambda$ = 543 nm using a modelocked supercontinuum fiber laser with a pulse duration of $\sim$ 5 ps and a repetition rate of 20 MHz. Excitation was performed through a microscope objective with numerical aperture (NA) of 0.63. The emission was collected near $\lambda$ = 620 nm using  a dichroic beamsplitter (Semrock FF593-Di02-25x36), a cutoff filter (Schott RG610), and a 14-nm band pass filter (Semrock FF01-620/14-25). The location of a single emitter is identified using a CCD camera and its time-resolved emission is measured by a single-photon avalanche photodiode with a time resolution of 50 ps.

We studied two types of random photonic media: slabs made from polystyrene (PS) and zinc oxide (ZnO) with thicknesses between $L$ = 4.0 and 16.5 $\mu$m, see Fig. \ref{fig1}(b). To probe the emission rates and thus the LDOS, we doped the samples with fluorescent polystyrene nanospheres with a diameter of 22(2) nm, an emission peak at 612 nm, and a quantum efficiency of 96(4) $\%$ (Duke Scientific red fluorescent nanospheres).
The fluorescent nanospheres are much smaller than the wavelength of light and contain about thirty dye molecules which all sense the LDOS at essentially the same position. Since the orientation of the molecules is random and the LDOS is orientation dependent \cite{perez2,VosKoenderink2009}, dye molecules inside one fluorescent nanosphere
emit at different rates. The fluorescent nanospheres show no photoblinking and are unaffected by the chemical environment since the dye molecules are protected from ambient oxygen and rigidly held in a polymer matrix. To fabricate PS random photonic media, a polydisperse suspension of spheres (Duke Scientific) doped with fluorescent nanospheres was deposited on a standard microscope cover slide, spread uniformly and allowed to dry. This inward-growing self-assembly method was previously demonstrated for photonic crystal fabrication \cite{Joan}. By preparing a polydisperse mixture of different spheres, crystallization was prevented. To fabricate ZnO random photonic media, a suspension of pigment (Aldrich ZnO $<$ 1 $\mu$m, 99 $\%$) doped with fluorescent nanospheres was sprayed on a cover slide \cite{elbert}. The random photonic media have a low areal density of fluorescent nanospheres ranging from 0.03 to 0.12 $\mu$m$^{-2}$. As non-scattering reference samples (with $1/k\ell$ = 0), we prepared fluorescent nanospheres in transparent polyvinyl alcohol polymer layers that were spincoated on cover slides and covered by Poly(methyl methacrylate) index matching layers \cite{epaps}.
The transport mean free paths for PS and ZnO random photonic media were determined by total transmission to be $\ell$ = 1.7(3) $\mu$m and $\ell$ = 0.82(12) $\mu$m  at $\lambda$ = 620 nm, respectively, yielding scattering strengths ($1/k\ell$) of 0.06 and 0.12, respectively \cite{fn_ell,elbert,epaps}. For time-resolved fluorescence, we measured a total of 96 single fluorescent nanospheres, namely 12 nanospheres in each of four PS and four ZnO samples. We also collected data from 12 single fluorescent nanospheres in the reference  polymer layer.

\begin{figure}[!ht]
\begin{center}
\includegraphics[width=3.25in]{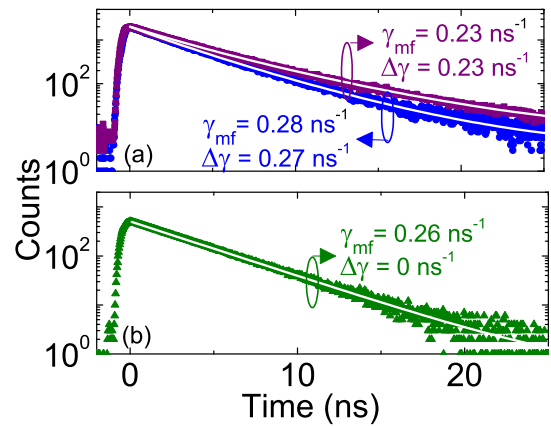}
\end{center}
\caption{\label{fig2}(color)(a) Time-resolved fluorescence from a single fluorescent nanosphere embedded in PS random photonic medium with a thickness of 7.4(5) $\mu$m. The slow (purple) and the fast (blue) decay curves were measured on two different single fluorescent nanospheres at the same depth $d$ = 3.3 $\mu$m. The white lines through the data are lognormal fits. The most-frequent decay rate \gmf ~and the width of the distribution \dlg ~are shown. (b) Time-resolved fluorescence of a single fluorescent nanosphere in a reference polymer sample (green) decays exponentially with a rate of 0.26 ns$^{-1}$.}
\end{figure}

A fluorescence image of a single fluorescent nanosphere embedded inside a PS random photonic medium is shown in Fig. \ref{fig1}(c). The image shows a diffuse spot of the fluorescent light. The background fluorescence of the host is negligible compared to the emission of the nanosphere. We determined the depth of the nanosphere by modeling the diffuse spot of fluorescent light with diffusion theory \cite{elbert,durian}. We conservatively estimate a depth accuracy of one mean free path. In the particular case of Fig.~\ref{fig2} we obtain $d$ = 3.3 $\pm$ 1.7$\mu$m. For our measurements, we selected only single fluorescent nanospheres that were isolated from other fluorescent nanospheres and situated at a depth of approximately half the sample thickness \cite{epaps}.

In Fig.~\ref{fig2}(a), we show the time-resolved fluorescence from two fluorescent nanospheres embedded in a PS random photonic medium at the same depth. The two nanospheres at different positions yield two different decay curves
demonstrating
the spatial fluctuation of the LDOS.
In Fig.~\ref{fig2}(b), the decay curve of a single fluorescent nanosphere in a non-scattering polymer layer is seen to be purely exponential. The time-dependent fluorescence of 12 single fluorescent nanospheres inside a polymer layer all show exponential decay with the same time constant of 0.26 ns$^{-1}$ to within only 1$\%$.
We conclude that non-exponential decay of a single fluorescent nanosphere is a characteristic of the LDOS fluctuations in random photonic media, and attribute the non-exponential decay to the strong dependence of emission rate on dipole orientation \cite{perez2}.

We quantitatively analyze the non-exponential time-resolved fluorescence curves with a distribution of emission rates, using the method that successfully describes time-resolved emission in ordered photonic media \cite{Ivan}. By choosing a lognormal distribution of emission rates we are ensured of physical positive rates with only one extra adjustable parameter as compared to the single-exponential model. The parameters of the distribution are the most-frequent emission rate \gmf, which is the peak of the distribution, and \dlg, which is its $1/e$ width. The mean emission rate \gmean ~is equal to $\gamma_{\mu} = \gamma_{mf}\exp\left(3/4\cdot{\left[\sinh^{-1}\left(\Delta{\gamma}/2\gamma_{mf}\right)\right]^{2}}\right)$ \cite{Ivan}. From the analysis of the emission curves of single probes inside PS samples in Fig. \ref{fig2}, the slowest emission curve is described by \gmean ~= $0.28$ ns$^{-1}$ while the fastest one has \gmean ~= $0.33$ ns$^{-1}$.

\begin{figure}[!ht]
\begin{center}
\includegraphics[width=3.25in]{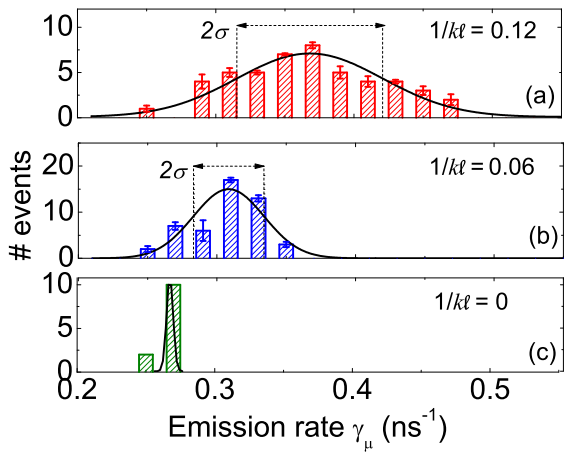}
\end{center}
\caption{\label{fig3}(color) Distribution of the emission rates for materials with different scattering strengths ($1/k\ell$). Histograms of the emission rates from fluorescent nanospheres embedded in three different materials: (a) ZnO, (b) PS, and (c) non-scattering polymer layer. The width of the distribution $2\sigma$ is indicated by the dotted lines. The full lines are Gaussian fits to the histograms.}
\end{figure}

Figs. \ref{fig3}(a) and (b) show histograms of emission rates for ZnO and PS random photonic media, respectively. As a measure of the width of the distributions we show twice the square root of the variance $ 2\sigma= 2\sqrt{{\rm Var} (\gamma_\mu)}$. The histogram for a non-scattering  reference  medium is shown in Fig. \ref{fig3}(c).
For all our samples, we verified that both the mean and the variance of the emission rate are independent of sample thickness.
The  distribution of emission rates is very narrow in the reference samples, showing that the fluorescent spheres have negligible intrinsic fluctuations. Interestingly, in moderately scattering PS samples ($1/k\ell$ =0.06), the width of distribution 2$\sigma$ = 0.05(1) ns$^{-1}$ is already much larger than in the reference samples. In strongly scattering ZnO samples ($1/k\ell$ =0.12), the distribution is twice as broad, 2$\sigma$ = 0.10(2) ns$^{-1}$.  We note in passing that the maximum of the distribution also shifts, probably due to a change in effective refractive index \cite{Schuurman}. The increase of the width tracks the increase of the scattering strength,  in qualitative agreement with the theoretical prediction that LDOS fluctuations increase with the scattering strength \cite{Mirlin,vanTiggelen}.

\begin{figure}[!ht]
\begin{center}
\includegraphics[width=3.25in]{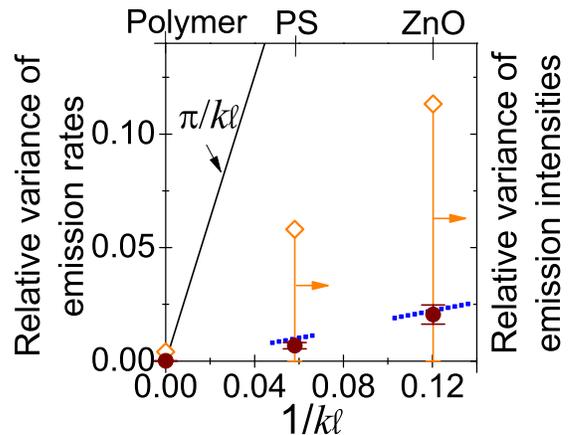}
\end{center}
\caption{\label{fig4}(color) (a) Relative variance of the mean decay rate $\sigma^{2}/\langle{\gamma_{\mu}}\rangle^{2}$ ~versus scattering strength $1/k\ell$. The measured values are shown as filled circles with error bars indicating the statistical uncertainty for $N$ = 48 measurements. The empty diamonds are the upper bound to $C_0$ determined from the fluctuations of the emitted intensity. The full and dotted lines show the LDOS variance for point-like scatterers \cite{vanTiggelen} and for the single-sphere-scatterer model, respectively.}
\end{figure}

The main result of our work shown in Fig. \ref{fig4} is the measurement of the relative variance of the emission rate $\sigma^{2}/\langle{\gamma_{\mu}}\rangle^{2}$ versus scattering strength ($1/k\ell$). For non-scattering samples ($1/k\ell = 0$) the relative variance vanishes ($\sigma^{2}/\langle{\gamma_{\mu}}\rangle^{2} < 10^{-4}$). For PS ($1/k\ell = 0.06$) and ZnO ($1/k\ell = 0.12$) samples, we find a relative variance of $0.007(2)$ and $0.020(4)$, respectively. The relative variance clearly increases with the scattering strength. Strikingly, the LDOS fluctuations predicted from the point-like scatterer model \cite{vanTiggelen} are about a factor twenty too high compared to our measurements. We attribute this difference to the fact that scalar wave and point scatterer approximations were used in Ref. \cite{vanTiggelen}, whereas in experiments the scatterers have finite size and the light waves have vector character. The vector character of light could have a strong effect as shown theoretically in Ref. \cite{perez2}.

As the point-like scatterer model does not provide an accurate description of the immediate surroundings of the emitter, we propose an alternative model to estimate the fluctuations of the emission rate, in which only one scatterer closest to the fluorescent nanosphere is considered. This is reasonable since in a dried sample, fluorescent nanospheres are always in contact with the surface of at least one scatterer. The calculation is done by adapting the analytical model for the spontaneous emission of an electric dipole located near a spherical surface \cite{chew}.  We calculate the emission rate of a dye molecule at all positions inside a fluorescent nanosphere in contact with the surface of a PS or ZnO spherical scatterer for all possible orientations of the molecule. The relative variance of the emission rate is obtained by averaging over the size distribution of the spherical scatterers \cite{epaps}. The calculated values are $0.010(3)$ and $0.022(5)$ for PS and ZnO samples, respectively, in good agreement with the experimental data (Fig. \ref{fig4})\cite{footnoteb}. This result confirms the hypothesis that the nearest scatterer dominates LDOS fluctuations in random photonic media \cite{Mirlin,vanTiggelen}.

It has been predicted that in infinite random media LDOS fluctuations are equivalent to $C_0$ fluctuations of emission intensity \cite{vanTiggelen}.  To probe $C_{0}$, we have measured the relative variance of the emission intensity of 20 individual fluorescent nanospheres inside 16.5 $\mu$m-thick PS and 13.3 $\mu$m-thick ZnO random photonic media. We note that the variance of the intensity provides an upper bound to $C_{0}$ since it consists of both residual speckle variance and the desired $C_{0}$ contribution \cite{epaps}. The intensity variance is therefore necessarily larger than $C_0$. The relative variance of emission intensity is shown in Fig. \ref{fig4}, it is indeed always higher than the relative variance of the emission rates. This observation means that our data are consistent with the hypothesis that the LDOS fluctuations and $C_0$ fluctuations are equivalent in our samples.

In conclusion, we observed fluctuations of the LDOS inside random photonic media through the spontaneous emission of individual fluorescent nanospheres embedded deep inside. Recently, an interesting study of emission rates near the interface of random materials has appeared \cite{sand}. The relative variance of LDOS fluctuations increases with scattering strength and agrees with a theoretical model.  We suggest that $C_0$ may also be responsible for the broad distribution of emission rates observed in photonic crystals \cite{Ivan}. In view of the overestimation of fluctuations by the point-like scatterer model, we anticipate that our observations will stimulate new
studies of the LDOS in random and partly ordered photonic media.

We thank Sanli Faez, Merel Leistikow and Ad Lagendijk for discussions, and Elbert van Putten and Hannie van den Broek for help with sample preparation.
This work is part of the research program of the "Stichting voor Fundamenteel Onderzoek der Materie" (FOM), which is financially supported by the
"Nederlandse Organisatie voor Wetenschappelijk Onderzoek" (NWO). APM and WLV are supported by NWO VIDI and VICI.


\begin{thebibliography}{99}
\bibitem{fn_ell} We define the scattering strength using the transport mean free path, which yields the most conservative value.
\end{thebibliography}
\end{document}